\documentclass[runningheads]{llncs}
\usepackage[T1]{fontenc}
\usepackage{graphicx,verbatim}

\usepackage{mathtools,amssymb}
\usepackage{microtype}      

\begin{document}
%
\title{PET Image Denoising via Text-Guided Diffusion: Integrating Anatomical Priors through Text Prompts}

\author{Boxiao Yu\inst{1} \and
Savas Ozdemir\inst{2} \and
Jiong Wu\inst{1} \and
Yizhou Chen\inst{3} \and
Ruogu Fang\inst{1} \and
Kuangyu Shi\inst{3} \and
Kuang Gong\inst{1}}

\authorrunning{B. Yu et al.}

\institute{J. Crayton Pruitt Family Department of Biomedical Engineering, University of Florida, Gainesville, FL, USA\and
Department of Radiology, University of Florida, Jacksonville, FL, USA\and
Department of Nuclear Medicine, University of Bern, Bern, Switzerland
}

\maketitle              
\begin{abstract}
Low-dose Positron Emission Tomography (PET) imaging presents a significant challenge due to increased noise and reduced image quality, which can compromise its diagnostic accuracy and clinical utility. Denoising diffusion probabilistic models (DDPMs) have demonstrated promising performance for PET image denoising. However, existing DDPM-based methods typically overlook valuable metadata such as patient demographics, anatomical information,  and scanning parameters, which should further enhance the denoising performance if considered. Recent advances in vision-language models (VLMs), particularly the pre-trained Contrastive Language–Image Pre-training (CLIP) model, have highlighted the potential of incorporating text-based information into visual tasks to improve downstream performance. In this preliminary study, we proposed a novel text-guided DDPM for PET image denoising that integrated anatomical priors through text prompts. Anatomical text descriptions were encoded using a pre-trained CLIP text encoder to extract semantic guidance, which was then incorporated into the diffusion process via the cross-attention mechanism. Evaluations based on paired $1/20$ low-dose and normal-dose $^{18}$F-FDG PET datasets demonstrated that the proposed method achieved better quantitative performance than conventional UNet and standard DDPM methods at both the whole-body and organ levels. These results underscored the potential of leveraging VLMs to integrate rich metadata into the diffusion framework to enhance the image quality of low-dose PET scans.


\end{abstract}

\section{Introduction}
\begin{figure}[htb]
\centering
\includegraphics[width=\textwidth]{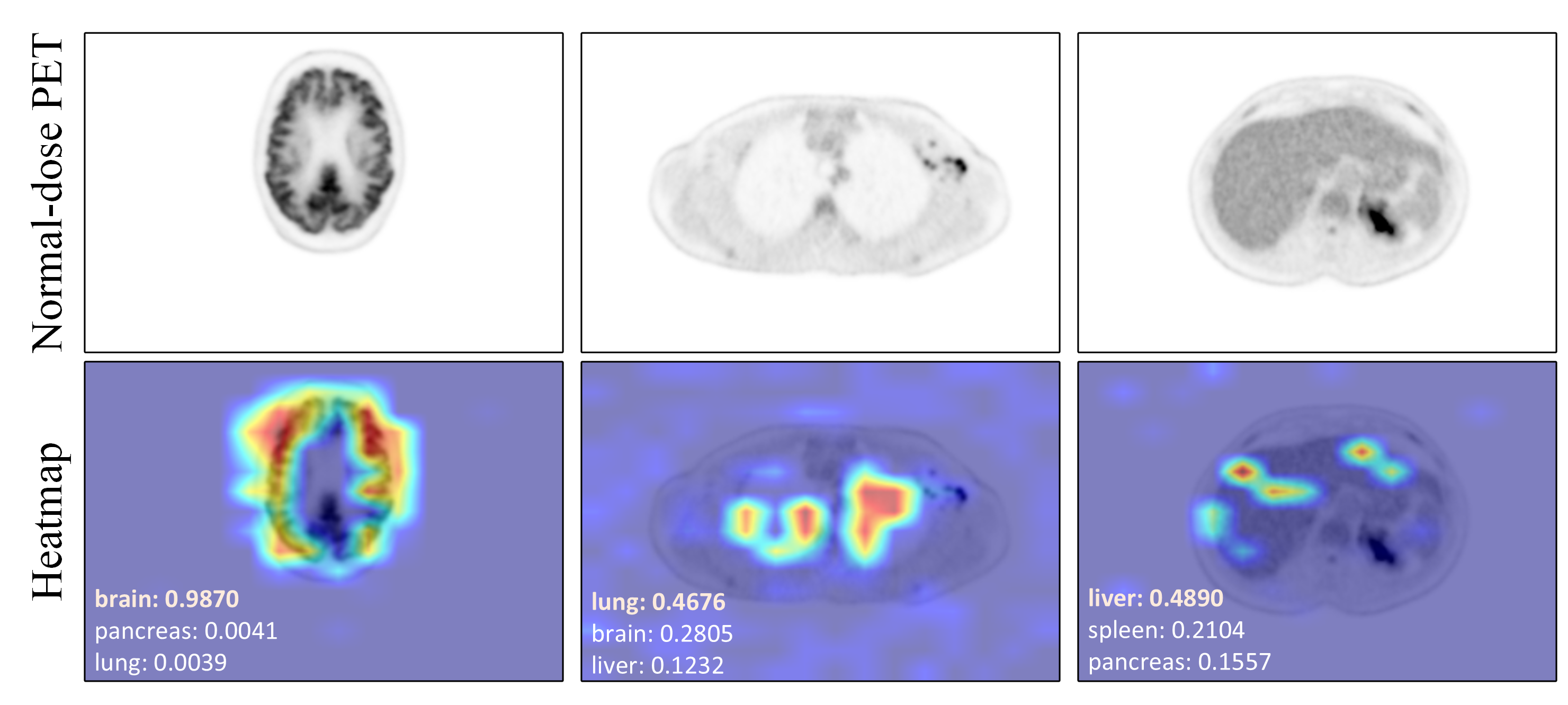}
\caption{Grad-CAM~\cite{selvaraju2017grad} results based on CLIP encoder, generated using axial PET slices and anatomical structures text prompts. The top figures show normal-dose PET images, while the bottom figures present heatmaps highlighting regions of interest, along with the top three anatomical text prompts that best match the PET image features. The highlighted regions correspond to critical anatomical structures of PET images, demonstrating that the CLIP text embeddings effectively capture relevant anatomical information.}
\label{Heatmap}
\end{figure}
Positron Emission Tomography (PET) is widely utilized in clinical and research settings for its high sensitivity and quantitative accuracy, making it indispensable in tumor detection, neurological diagnosis, and metabolic function evaluation~\cite{barthel2015pet,ming2020progress}. Considering the radiation risks caused by radiotracer injections, faster or low-dose PET imaging is always desired. This trade-off results in limited counts received during acquisition, leading to higher noise and lower signal-to-noise ratio (SNR), potentially compromising its diagnostic and quantitative accuracy. Obtaining high-quality PET images from low-dose scanning protocols remains a key clinical challenge.

In recent years, deep learning methods have revolutionized the image processing field. The diffusion model~\cite{dhariwal2021diffusion,ho2020denoising}, an advanced distribution learning-based method, has shown promising performance in various computer vision tasks~\cite{jiang2024fast,lee2025guaranteed,luo2023controlling}. Compared with traditional methods such as Convolutional Neural Networks (CNNs)~\cite{gong2018pet,zhang2024mango,hashimoto2019dynamic} and Generative Adversarial Networks (GANs)~\cite{wang20183d,wang20183d,fu2023aigan} that directly map from a source to a target domain, diffusion models iteratively reconstruct the target distribution from Gaussian distribution via a two-stage diffusion process. This approach has demonstrated greater stability and improved image fidelity. Existing literatures\cite{yu2024pet,yu2025robust,jiang2023pet,shen2023pet} have shown its effectiveness in PET image denoising.

Nevertheless, existing diffusion-based PET enhancement methods typically overlook valuable clinical metadata, such as patient demographics, slice information, and scanning protocols, which contain patient-specific details and other factors that potentially affect the image quality. Incorporating such metadata information might further improve image quality. Some previous CNN-based studies~\cite{lesonen2024anatomy,yang2024anatomically,xie2020anatomy} have attempted to integrate complementary features from CT or MR images into the PET reconstruction framework; however, the limited availability of paired PET/CT or PET/MR data restricts their application. 
Alternatively, Luo~et~al.~\cite{luo2025weight} proposed transforming clinical anatomical information into structural attributes via one-hot encoding to adjust the weights of different body regions during PET image generation. However, one-hot encoding fails to capture the semantic relationships between structures, thereby limiting its capacity to effectively align with the complex features present in PET images.

Recent advances in large-scale vision-language models (VLMs)~\cite{liu2023visual,li2023llava,li2022blip} have revealed great potential in extracting robust visual and textual representations for downstream tasks. The Contrastive Language–Image Pre-training (CLIP) model~\cite{radford2021learning}, for example, learns aligned multimodal features from noisy image-text pairs via contrastive learning. As shown in Fig.~\ref{Heatmap}, CLIP-generated text embeddings based on anatomical descriptions (e.g., "brain", "lung", "liver") can effectively match corresponding PET image features, demonstrating their ability to identify and focus on critical anatomical regions in PET images. In this work, we integrated anatomical text prompts into a diffusion-based PET image denoising framework. By leveraging semantic guidance from pre-trained CLIP through a cross-attention mechanism, our approach explicitly guided the model to focus on key anatomical structures, thus enhancing the accuracy of organ-intensity restoration during the denoising process. Experimental results based on clinical total-body PET datasets demonstrated that the proposed method outperformed traditional UNet and DDPM methods in both global and organ-based quantitative accuracy.

\section{Methodology}
\subsection{Conditional DDPM for PET Image Denoising} 
Denoising Diffusion Probabilistic Model (DDPM)~\cite{ho2020denoising} provides a powerful framework for image generation by modeling a gradual denoising process. In this work, we adopted DDPM for PET image denoising as a baseline method, leveraging their ability to iteratively refine noise inputs into high-quality outputs. The process consisted of two phases: a forward diffusion process that gradually corrupted the input image with Gaussian noise, and a reverse diffusion process that reconstructed the clean image from pure Gaussian noise.

The forward diffusion process transformed a high-quality normal-dose PET image $\mathbf{x}_0$ into a noisy version $\mathbf{x}_T$ over $T$ time steps, following a predefined variance schedule $\{\beta_t\}^T_{t=1}$, which is defined as a Markov chain:
\begin{align}
q(\mathbf{x}_{1:T} | \mathbf{x}_0) &= \prod_{t=1}^T q(\mathbf{x}_t | \mathbf{x}_{t-1} ), \qquad 
q(\mathbf{x}_t|\mathbf{x}_{t-1}) = \mathcal{N}(\mathbf{x}_t;\sqrt{1-\beta_t}\mathbf{x}_{t-1},\beta_t \mathbf{I}).
\end{align}
By introducing $\alpha_t = 1-\beta_t$ and $\bar\alpha_t = \prod_{s=1}^t \alpha_s$, the intermediate noisy image $\mathbf{x}_t$ at any time step can be directly sampled from $\mathbf{x}_0$:
\begin{align}
q(\mathbf{x}_t|\mathbf{x}_0) = \mathcal{N}(\mathbf{x}_t; \sqrt{\bar{\alpha}_t}\mathbf{x}_0, (1 - \bar{\alpha}_t)\mathbf{I}).
\end{align}

The reverse diffusion process aimed to recover the clean image $\mathbf{x}_0$ from the noisy observation $\mathbf{x}_T \sim \mathcal{N}(0, \mathbf{I})$, which was achieved by learning a neural network $p_{\boldsymbol{\theta}}$ that approximated the conditional distribution $q(\mathbf{x}_{t-1}|\mathbf{x}_t)$:
\begin{align}
  p_\theta(\mathbf{x}_{t-1}|\mathbf{x}_t, \mathbf{y}) = \mathcal{N}(\mathbf{x}_{t-1}; \boldsymbol{\mu}_\theta(\mathbf{x}_t, \mathbf{y}, t), \sigma_t^2 \mathbf{I}),
\end{align}
where $\mathbf{y}$ represents the observed low-quality PET image used as a conditional input to guide the denoising process. Instead of directly predicting the mean $\boldsymbol{\mu}_{\boldsymbol{\theta}}$, the model was trained to estimate the noise $\boldsymbol{\epsilon}_{\boldsymbol{\theta}}$ added during the forward process based on the UNet~\cite{ronneberger2015u} backbone. This simplified the learning objective, which minimized the difference between the predicted and true noise as
\begin{align} 
    L(\boldsymbol{\theta}) = \mathbb{E}_{\mathbf{x}_0, \mathbf{y}, t, \boldsymbol{\epsilon}} \left[ \left\| \boldsymbol{\epsilon} - \boldsymbol{\epsilon}_{\boldsymbol{\theta}}\left(\mathbf{x}_t,  \mathbf{y}, t\right) \right\|^2 \right] . 
\end{align}
Consequently, the iterative denoising process was governed by:
\begin{align}
\mathbf{x}_{t-1} = \frac{1}{\sqrt{\alpha_t}}\left( \mathbf{x}_t - \frac{\beta_t}{\sqrt{1-\bar\alpha_t}} \boldsymbol{\epsilon}_{\boldsymbol{\theta}}(\mathbf{x}_t, \mathbf{y}, t) \right) + \sigma_t \mathbf{z},\quad \text{where}~\mathbf{z} \sim \mathcal{N}(0, \mathbf{I}). \label{final_calculation}
\end{align}
This framework established the baseline for our PET image denoising method. In the following section, we further extended this framework by incorporating anatomical information as text prompts.

\subsection{Integrating Anatomical Priors as Text Prompts in DDPM}
\begin{figure}[htb]
\includegraphics[width=\textwidth]{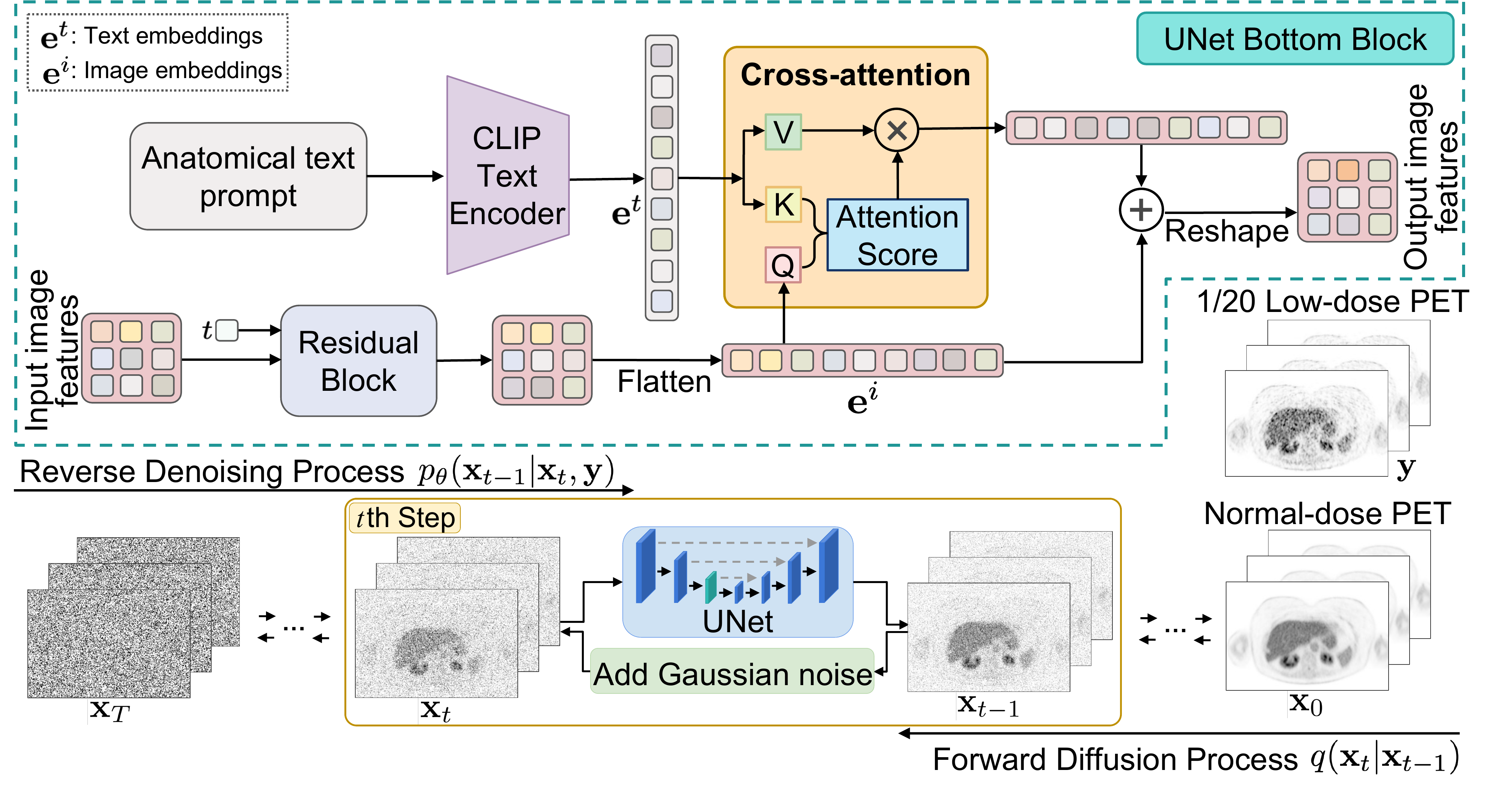}
\caption{Diagram of the proposed text-guided DDPM framework.} \label{fig_model}
\end{figure}
 
To further improve the quality of denoised PET images, we proposed a novel text-guided DDPM that incorporated domain-specific anatomical priors into the denoising process. Overall, textual prompts describing the major anatomical structures of each axial PET slice (e.g., “Axial slice including spleen, liver, stomach, aorta, vertebrae T10, ...”) were injected into the diffusion process, and a cross-attention mechanism was introduced to learn semantic guidance from the pre-trained CLIP encoder. 

Fig.~\ref{fig_model} illustrates the diagram of the proposed text-guided DDPM. Specifically, the anatomical textual prompts were first encoded by the CLIP text encoder into high-dimensional anatomical text embeddings $\mathbf{e}^t$, which captured organ- and structure-level semantics for each axial slice. 
During training, these embeddings were then fed into the bottom blocks of the UNet via cross-attention, where the query features extracted from PET images and text-derived keys computed the attention score to determine how the anatomical information modulated the image.
Consequently, the score function of the diffusion network was modified to $\boldsymbol{\epsilon}_\theta(\mathbf{x}_t,  \mathbf{y}, t, \mathbf{e}^t)$, making the denoising process explicitly aware of the underlying anatomical context.
Unlike conventional DDPMs that rely solely on paired images, the proposed model leveraged CLIP’s ability to learn robust text-image alignments. By embedding high-level anatomical cues into the feature space, the network was prompted to focus on the critical anatomical regions, enabling the UNet to adaptively refine its predictions based on the specific anatomical context of each PET slice. 
\section{Experiments and Results}
\begin{figure}[htb]
\includegraphics[width=\textwidth]{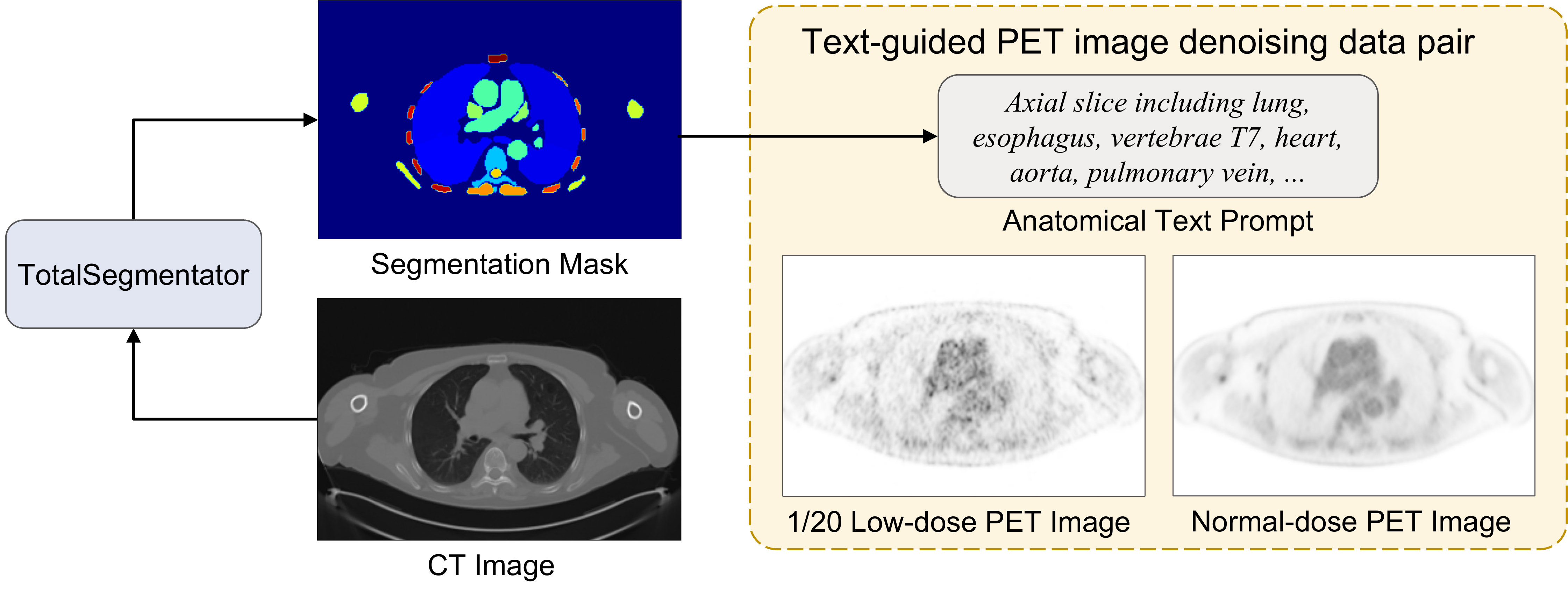}
\caption{Overview of the dataset construction for text-guided PET image denoising.} \label{fig_dataset}
\end{figure}
\subsection{Datasets}
We constructed a text-guided PET image denoising dataset that integrated anatomical information. It comprised  paired $1/20$ low-dose and normal-dose PET images together with corresponding anatomical text descriptions for each axial slice. The $^{18}$F-FDG PET images were obtained from the Siemens Biograph Vision Quadra PET/CT scanner and reconstructed using OSEM ($4$ iterations, $5$ subsets) with time-of-flight (TOF) and point spread function (PSF) modeling. A $2 \ \mathrm{mm}$ Gaussian filter was applied~\cite{alberts2021clinical}. The resulting voxel size was $1.65 \times 1.65 \times 1.65 \, \text{mm}^3$. All data were acquired in list mode, enabling rebinding to simulate various dose levels. The $1/20$ low-dose PET images were generated through subsampling at $1/20$ of the original dose. The image intensity was expressed in Standardized Uptake Value (SUV). A total of $115$ whole-body PET scans were utilized, randomly divided into $90$ for training, $5$ for validation, and $20$ for testing. To improve computational efficiency, parts of background regions were cropped, yielding image matrices of $192 \times 288 \times 520$ corresponding to the coronal, sagittal, and axial dimensions, respectively.

Anatomical text prompts were derived from the corresponding CT images. Using the TotalSegmentator~\cite{wasserthal2023totalsegmentator} toolkit, which was based upon nnUNet~\cite{isensee2021nnu} and trained on CT and MR images from diverse scanners, institutions, and protocols, were adopted to obtain segmentation masks for $117$ major anatomical categories. These segmentation maps were processed to generate a mask for each axial PET slice. Unique anatomical categories present in each mask were identified to create text prompts listing the key anatomical structures for that slice. These text prompts were paired with the $1/20$ low-dose and normal-dose PET images and served as guidance in both training and testing of the text-guided DDPM. Fig.~\ref{fig_dataset} outlines the dataset preparation process.

\subsection{Implementation Details}

Textual anatomical embeddings were obtained by feeding the generated text prompts into the CLIP text encoder, with token sequences truncated to a maximum length of $77$ to align with CLIP’s input dimension. The utilized pre-trained CLIP model employs a ViT-B/32 Transformer as its image encoder and a masked self-attention Transformer as its text encoder.

Inspired by the work of Gong~et~al.~\cite{gong2023pet}, two adjacent axial slices were jointly fed into the network during training and testing to mitigate axial artifacts. All implementations were developed in PyTorch, with a learning rate of $1 \times 10^{-4}$ and the mixed-precision training strategy was employed to enhance computational efficiency. In both the forward and reverse diffusion processes, the time step $T$ was set to $1000$, with the variance schedule $\beta_t$ linearly increasing from $1 \times 10^{-4}$ to $0.02$. The total training batch size was set to $48$, and distributed training was performed on $6$ NVIDIA A100 GPUs. The entire training process required approximately two days, with an average testing time of $24$ minutes per dataset. For comparison, reference models including UNet and the original DDPM were trained on the same dataset under identical conditions.
\begin{figure}[htb]
\includegraphics[width=\textwidth]{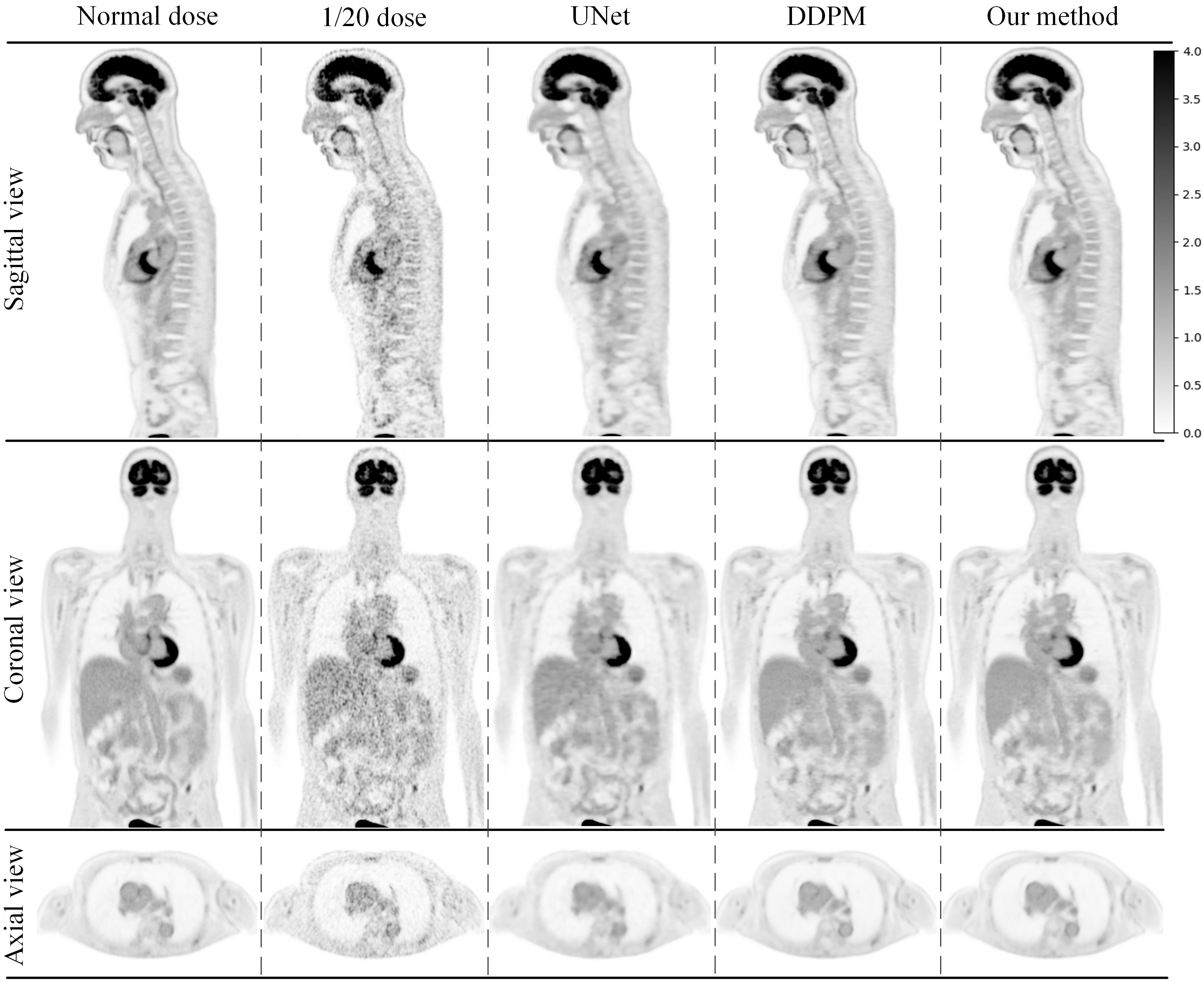}
\caption{Qualitative comparison of PET images in sagittal (top), coronal (middle), and axial (bottom) views. From left to right: normal-dose PET image, $1/20$ low-dose PET image (as input), and generated denoised results from UNet, DDPM, and our proposed method. The intensity scale on the right indicates SUV values. } \label{visual}
\end{figure}

\subsection{Evaluation Metrics}

To assess the quantitative performance of different denoising methods, two widely-used quantitative metrics: Peak Signal-to-Noise Ratio (PSNR) and Structural Similarity Index Measure (SSIM) were adopted, using the normal-dose PET images as the ground truth. All evaluations were initially conducted over the entire image, and were further computed for individual organs based on segmentation masks to assess the local denoising quality. Finally, Wilcoxon signed-rank tests were performed on the PSNR and SSIM scores to statistically validate the performance differences of different methods.

\subsection{Results}

\begin{figure}[htb]
\includegraphics[width=\textwidth]{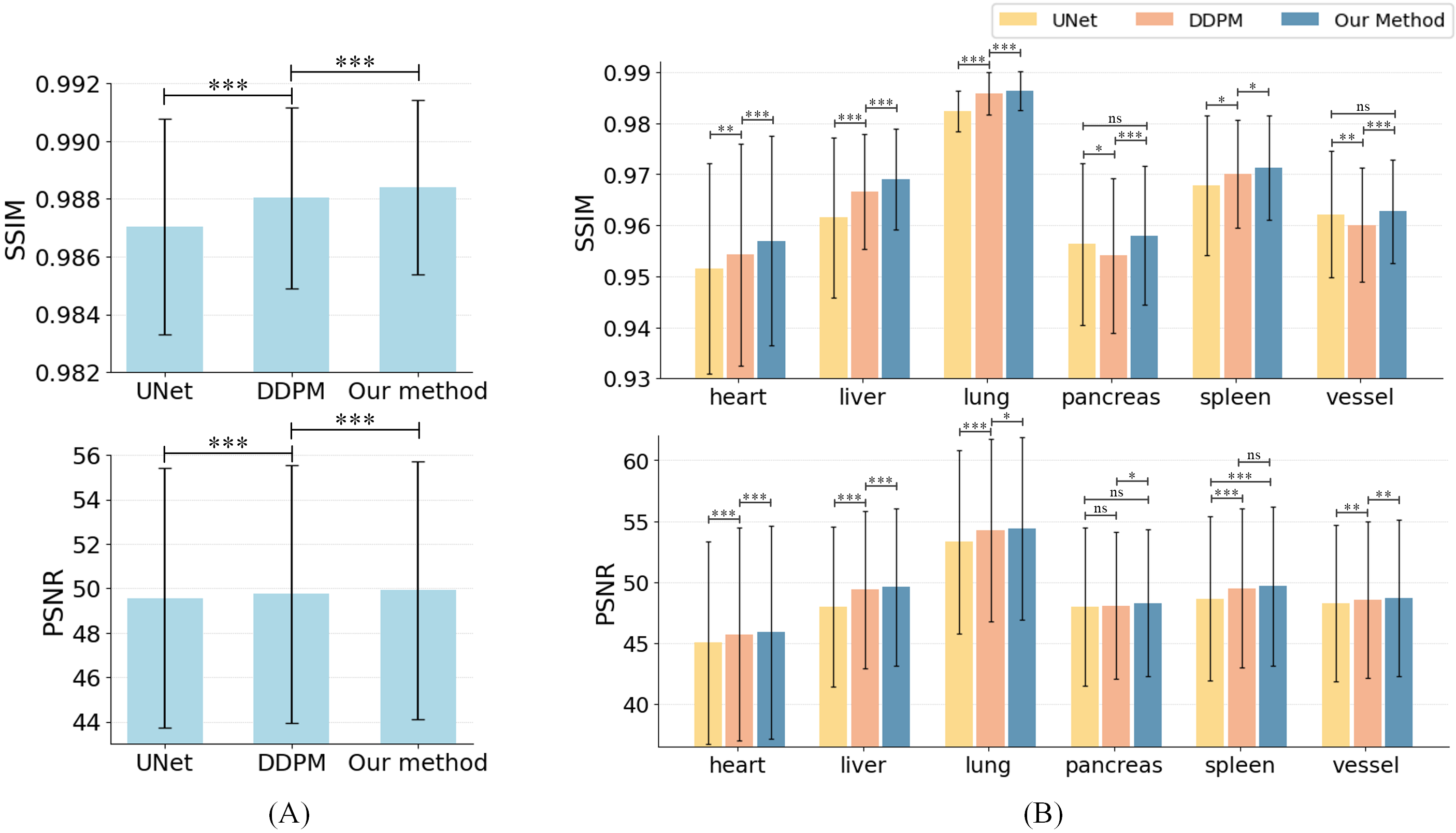}
\caption{Quantitative comparison of UNet, DDPM, and the proposed method in terms of SSIM and PSNR based on 20 $1/20$ low-dose test datasets. (A) Overall performance across the entire image. (B) Organ-specific results for the heart, liver, lung, pancreas, spleen, and vessels. ***, **, *, ns located at the top of the bar plot represents p-value\textless $0.001$, p-value\textless $0.01$, p-value\textless $0.05$, p-value$\geq$$0.05$, respectively.} \label{quant}
\end{figure}

Fig.~\ref{visual} shows a qualitative comparison of different methods. The $1/20$ dose images exhibited notable noise and low SNR, while all three denoising methods restored more anatomical details. However, the UNet outputs appeared over-smoothed, while both DDPM and the proposed method produced more realistic images closer to the normal-dose PET reference. Notably, the proposed method reduced artifacts around organs and provided sharper edge contours.

Fig.~\ref{quant} shows the quantitative comparison between UNet, DDPM, and our method on the entire body and individual organs. Overall, the proposed model achieved higher SSIM and PSNR than the other methods, indicating better denoising performance. Organ-specific metrics demonstrated consistent improvements in regions such as the heart, liver, and spleen, where anatomical clarity and quantitative accuracy were crucial for diagnostic assessments. Although the magnitude of improvement varied across different organs, performance gains are statistically significant, highlighting the robustness of the proposed text-guided approach.

\section{Conclusion}

In this work, we proposed a novel text-guided diffusion model for PET image denoising that integrated anatomical priors through text prompts. By leveraging semantic information encoded via the pre-trained CLIP text encoder, the proposed method achieved better performance on both whole-body and organ-specific evaluations. These preliminary results underscored the potential of incorporating multi-modal semantic guidance through vision-language models to enhance PET image quality. Our future work will focus on fine-tuning the CLIP text encoder for PET  domain and integrating additional metadata information.

\section{Acknowledgments}
This work was supported by NIH grants R01EB034692 and R01AG078250.
%
%
%
\bibliographystyle{splncs04}
\bibliography{mybibliography}

\end{document}